\documentclass[journal=apchd5,manuscript=article]{achemso}
\usepackage{graphicx}
\usepackage{amsmath}
\usepackage{array}
\usepackage{booktabs}
\usepackage{hypdoc}
\usepackage{listings}
\usepackage{lmodern}
\usepackage{mathpazo}
\usepackage{microtype}
\usepackage[T1]{fontenc}
\usepackage{hyperref}
\hypersetup{
    colorlinks=true,
    linkcolor=blue,
    filecolor=black,      
    urlcolor=blue,
    allcolors=blue,
}
\mciteErrorOnUnknownfalse
\title{Electro-mechanically tunable, waveguide-coupled photonic-crystal cavities with embedded quantum dots}
\author{L. A. F. Brunswick}
\affiliation{School of Mathematical and Physical Sciences, University of Sheffield, UK}%
\email{l.brunswick@sheffield.ac.uk}
\author{L. Hallacy}
\affiliation{School of Mathematical and Physical Sciences, University of Sheffield, UK}
\author{R. Dost}
\affiliation{School of Mathematical and Physical Sciences, University of Sheffield, UK}
\author{E. Clarke}
\affiliation{School of Electrical and Electronic Engineering, University of Sheffield, UK}%
\author{M. S. Skolnick}
\affiliation{School of Mathematical and Physical Sciences, University of Sheffield, UK}
\author{L. R. Wilson}
\affiliation{School of Mathematical and Physical Sciences, University of Sheffield, UK}
\keywords{Quantum dots, Nanophotonics, Purcell effect, Photonic resonators, Micro-electromechanical systems, Full tunablility}
\begin{document}
\begin{abstract}
On-chip micro-cavities with embedded quantum emitters provide an excellent platform for high-performance quantum technologies. A major difficulty for such devices is overcoming the detrimental effects of fluctuations in the device dimensions caused by the limitations of the fabrication processes. We present a fully tunable system based on a 1D photonic-crystal cavity with an embedded quantum dot, which enables tuning of both the quantum dot emission energy and the cavity mode wavelength. A micro-electromechanical cantilever is used to tune the cavity mode wavelength via index modulation and the quantum-confined Stark effect is used to tune the quantum dot emission energy, mitigating the effect of fabrication imperfections. To demonstrate the operation of the device, a maximum, voltage-controllable cavity tuning range of $\Delta \lambda = 1.8$ nm is observed. This signal is measured at the end of a bus waveguide which side-couples to the cavity, enabling the coupling of multiple cavities to a common waveguide, a key requirement for scale-up in these systems. Additionally, a quantum dot is tuned into resonance with the cavity mode, exhibiting an enhanced emission rate with a resolution limited Purcell factor of $F_P = 3.5$.

\end{abstract}
\section{Introduction}
 Semiconductor quantum dots (QDs) embedded in nanostructures have been an active field of research for the past two decades, during which significant progress has been made towards the realisation of quantum technologies based on this architecture. Demonstrations of key properties such as transform-limited emitter linewidths~\cite{transformlinewidth,airinterfacedegrade}, highly indistinguishable single-photon emission~\cite{lowg2,idsphotons}, strong light-matter interactions~\cite{H1singlephoton,gwppurcell,PHCWGpurcell,wvgnonlin,photoniccrystalcav} and highly efficient, deterministic single-photon generation~\cite{opencavity,QDembedded} have established QDs as a leading single-photon source for quantum technology applications. \\
Embedding QDs in optical resonators such as micro-cavities enhances many of their favourable properties due to the enhanced light-matter interaction that arises from coupling between the QD and the cavity when they are in resonance. In the weak-coupling regime, this enhancement is known as the Purcell effect~\cite{purcelleffect} which manifests itself as a reduction in the radiative lifetime of the QD~\cite{rateenhancement}. This phenomenon not only increases the emission rate of single-photons from the QD, but also reduces the sensitivity to losses caused by decoherence or non-radiative recombination~\cite{OChigheff}, which is critical to achieve high performance devices based on single quantum emitters.\\
To enhance the light-matter interactions of a QD in a micro-cavity, the QD emission must be in resonance with the optical mode of the cavity. This requirement poses a significant challenge for semiconductor QD systems as, due to the random nature of QD growth and fluctuations in fabrication~\cite{disordertheory,10.1063/1.4788709}, it is unlikely that a QD that is located in a cavity will be on resonance with the mode of the cavity. To grow QDs with high-quality optical characteristics the Stranski–Krastanov method is often used, which nucleates QDs in random positions on the substrate, and of random sizes, resulting in a distribution of emission energies~\cite{SKgrowth}. Furthermore, fabrication imperfections can cause nanostructure dimensions to deviate from their design, leading to significant detuning of the optical modes from their intended wavelength. Consequently, a method to independently control both the QD emission energy and the cavity resonance is highly desirable to maximise the yield of these devices.\\
The tuning of QD emission energies is a well-established technique in the field and can be achieved through several different, complementary methods such as electrical tuning via the quantum-confined Stark effect~\cite{QCSEmax}, strain tuning,~\cite{straintheory,strainexp,strainuniax,HfOtuning} magnetic field tuning~\cite{magnettune} and temperature tuning~\cite{thermaltune}. When considering cavity tuning methods, they can be split broadly into two categories: material or mode perturbation. Material perturbation methods influence the whole cavity/emitter system by altering the physical properties of the structure. Examples of material perturbation tuning methods include thermal~\cite{thermaltuning1,thermaltuning2}, electro-optical~\cite{MMIphase}, acoustic~\cite{acoustictuning} and strain~\cite{H1latticestrain,H1latticestretch}. However, these approaches cannot independently tune the QDs relative to the cavity mode.
Mode perturbation tuning methods modulate the effective refractive index of the cavity mode by using an electromechanical device to control the displacement of a dielectric material within the evanescent field of the cavity mode. This approach has the advantage of preserving the optical properties of the QD while enabling independent control of the cavity mode wavelength. Moreover, such an interaction is inherently local to a single cavity, a key requirement for a scaleable cavity emitter system. Such devices have seen success in silicon ~\cite{1DPhCCtriplebeamtuning,PhCCAFMtuning,1DPhCCsinglebeamtuning,cavtune3,Lin:15,8456716,8820141}, diamond - AlN hybrid ~\cite{tunecav1} and, more recently, in GaAs ~\cite{1DPhCCFiltering,cavtune2,cavtune4,midoloMich} systems. 
Actuation can occur in either in-plane or out-of-plane geometries. In-plane actuation requires intricate comb drives, which are complex to fabricate, occupy a large footprint and typically require moderate voltages on the order of 10 V to operate~\cite{midoloMZI}. Out-of-plane systems on the other hand, rely on the electrostatic attraction between layers of the wafer to displace the material, simplifying fabrication and reducing device size.\\
In the present work we embed InAs QDs in a 1D-photonic-crystal cavity (PhCC) where the electric field maximum is located in the high index material. Additionally, a perturbing beam is mounted on a singly clamped cantilever to tune the cavity. This geometry brings two key benefits relative to previous reports in GaAs systems. Firstly, 1D-PhCCs possess intrinsically low mode volumes, on the order of $\frac{1}{2} (\frac{\lambda}{n})^{3}$~\cite{10.1063/1.3107263}, providing a favourable environment for strong light-matter interactions. Secondly, the evanescent field of the cavity mode is easily accessible in a side-coupled geometry from both sides of the cavity~\cite{kspacecoupling} allowing for the possibility of simultaneous cavity-waveguide coupling and cavity mode tuning. Such a device could enable waveguide-mediated, cavity-cavity coupling with the integration of single quantum emitters. Such systems have recently been shown to have applications as high-efficiency, high-fidelity broadband single-photon switches~\cite{singlephotonswitch}, an important component in scaleable photonic circuits. 

Our electromechanical approach allows precise, voltage-controllable tuning of the cavity wavelength independent of the QD emission energy. Here, we present the design and optimisation of the cavity and cantilever, with experimental characterisation of the cavity tuning behaviour. Finally, we demonstrate the capabilities of the device by measuring the cavity mode tuning signal from the device output coupler, highlighting the waveguide coupling capabilities of our device. Additionally, we tune a QD through the cavity mode and measuring the resulting Purcell enhancement. 

\begin{figure}[t!]
\centering
\includegraphics[width=1\textwidth]{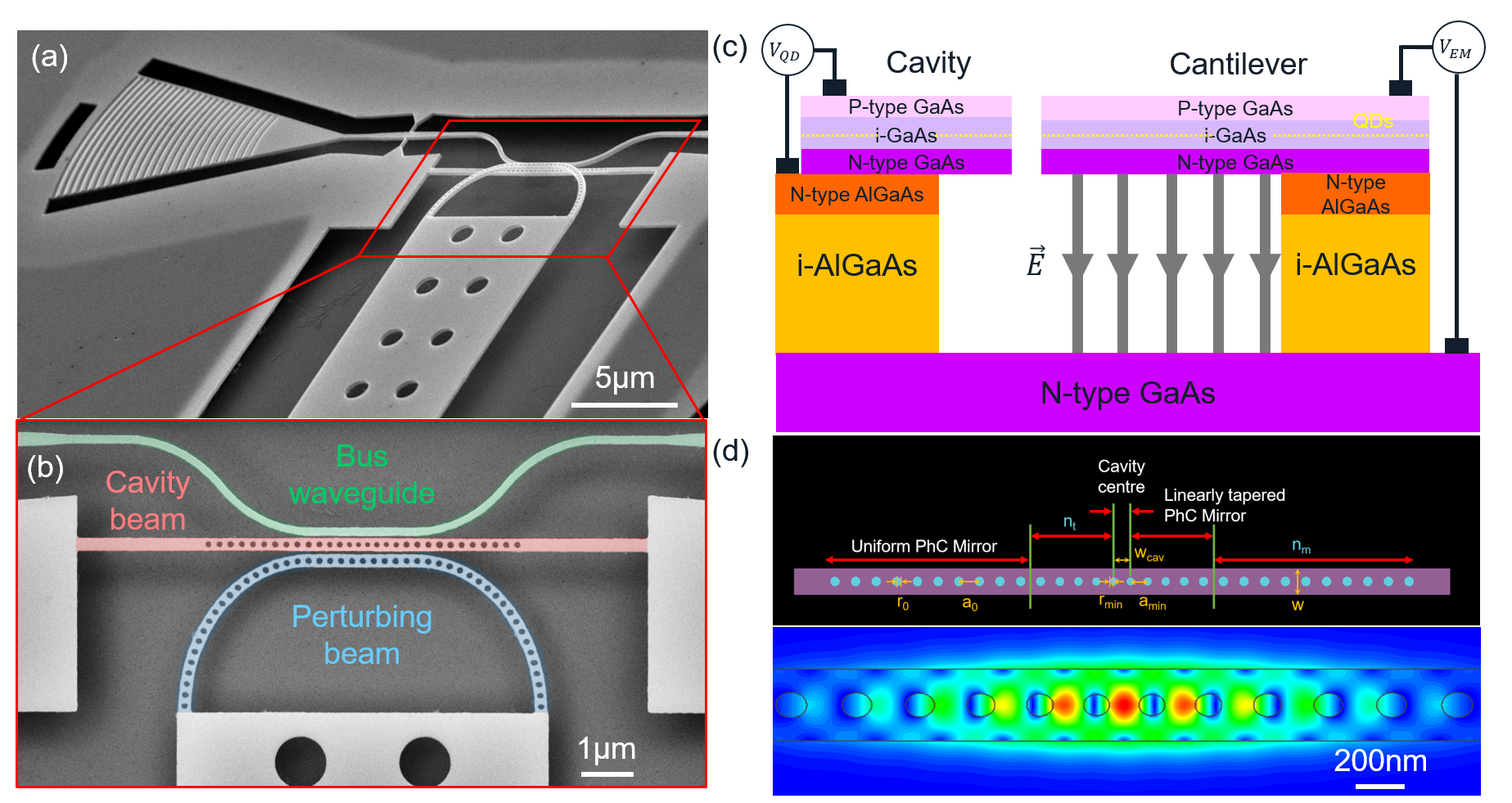}
\caption{(a) SEM image of an electro-mechanically tunable, waveguide-coupled cavity device. (b) False-colour SEM image of the cavity-waveguide interface region. The green, red, and blue sections denote the bus waveguide, cavity beam, and perturbing beam, respectively. (c) Schematic of the wafer and diode structures in the device. Applied voltage $V_{QD}$ tunes the QD transition energy and $V_{EM}$ moves the cantilever towards the substrate via an electrostatic force. (d) Top: Schematic of the 1D-PhCC design, optimised via FDTD simulations. Bottom: Simulated electric field profile of the cavity’s fundamental resonance.}
\label{fig:1}
\end{figure}
\section{Results}
\subsection{Device Design}
\autoref{fig:1} details the design and operational principle of our device. An SEM image of the device is shown in \autoref{fig:1} (a). Each device consists of a singly clamped cantilever terminated by a 1D photonic-crystal (perturbing beam) which, at rest, lies in close proximity to a 1D-PhCC. A bus waveguide is present on the opposite side of the cavity to enable side-coupling to the cavity (see \autoref{fig:1} (b)). This waveguide is terminated by a shallow-etched grating output coupler optimised for the QD emission wavelength (900 - 920 nm). The structures are fabricated in a 170 nm thick GaAs membrane containing self-assembled InAs QDs. The top and bottom layer of the membrane are p and n doped, respectively, enabling the formation of a p-i-n diode across the QD layer and allowing $V_{QD}$ to be applied across the QDs. Beneath the membrane, an AlGaAs sacrificial layer is selectively etched after the structures are patterned into the membrane to create free-standing devices. The full wafer structure is shown in \autoref{fig:1} (c). Importantly the substrate, directly under the AlGaAs sacrificial layer, is n-doped enabling the creation of a p-i-n-i-n diode across the structure. This allows $V_{EM}$ to be applied between the cantilevers and the substrate. 
To isolate the tuning of QDs from the cantilevers, a trench is etched through the p-layer at the base of the cantilever to break the electrical continuity between the cantilevers and the cavities. A cantilever size of 35.0 $\times$ 7.5 $\mu$m is chosen to give the best compromise between a low actuation voltage and resistance to drooping when the sacrificial layer is removed.\\A schematic and electric field profile of the 1D-PhCC cavity are shown in \autoref{fig:1} (d). The cavity beam contains three regions: the cavity centre with width $c_w$, a linear taper with minimum pitch $a_{min}$, minimum air hole radius $r_{min}$ and number of periods $n_t$, and a uniform mirror with pitch $a$, air hole radius $r_0$ and number of periods $n_m$. The thickness of the nanobeam $t$ is fixed by the wafer parameters and the width is chosen to be $w = 280$ nm to support single mode propagation. Each parameter was optimised using FDTD simulations to conduct parameter sweeps for fixed nanobeam dimensions. The optimised cavity design exhibited a Q-factor of $Q \sim 5\times 10^{6}$ without losses and a modal volume of $V_{m} = 0.48(\frac{\lambda}{n})^3$ (where $n = 3.4$). The design of the taper region is critical to achieving the optimal Q-factor as it acts to smooth the interface between the cavity and mirror modes which introduces scattering losses~\cite{PHCCtaper,modemismatch}.  In practice, as we discuss later in \autoref{sec:2.2}, the cavity Q-factors are reduced into the range of a few thousand due to the presence of inadvertent losses. In our design, we employ a linear taper where the ratio between the pitch and radius is constant i.e. $\frac{a_0}{r_0} = \frac{a_{min}}{r_{min}}$. The optimal value obtained from our simulations of $a_{min} = 0.84a_0$ is consistent with previous studies in the literature~\cite{bracher1DPhCC}.\\The perturbing beam design was also optimised to minimise scattering losses associated with beam displacement in the evanescent cavity field while achieving a large tuning range. Similarly, the bus waveguide design was optimised to balance between the cavity-waveguide coupling strength, and the cavity Q-factor. This was achieved by carefully adjusting the separation between the cavity and bus waveguide and modifying the waveguide width to optimise the spatial and k-space overlap between the cavity and waveguide modes (see \autoref{sec:buswgcoupling} for more detail).
\\ 
\begin{figure}[t!]
\centering
\includegraphics[width=1\textwidth]{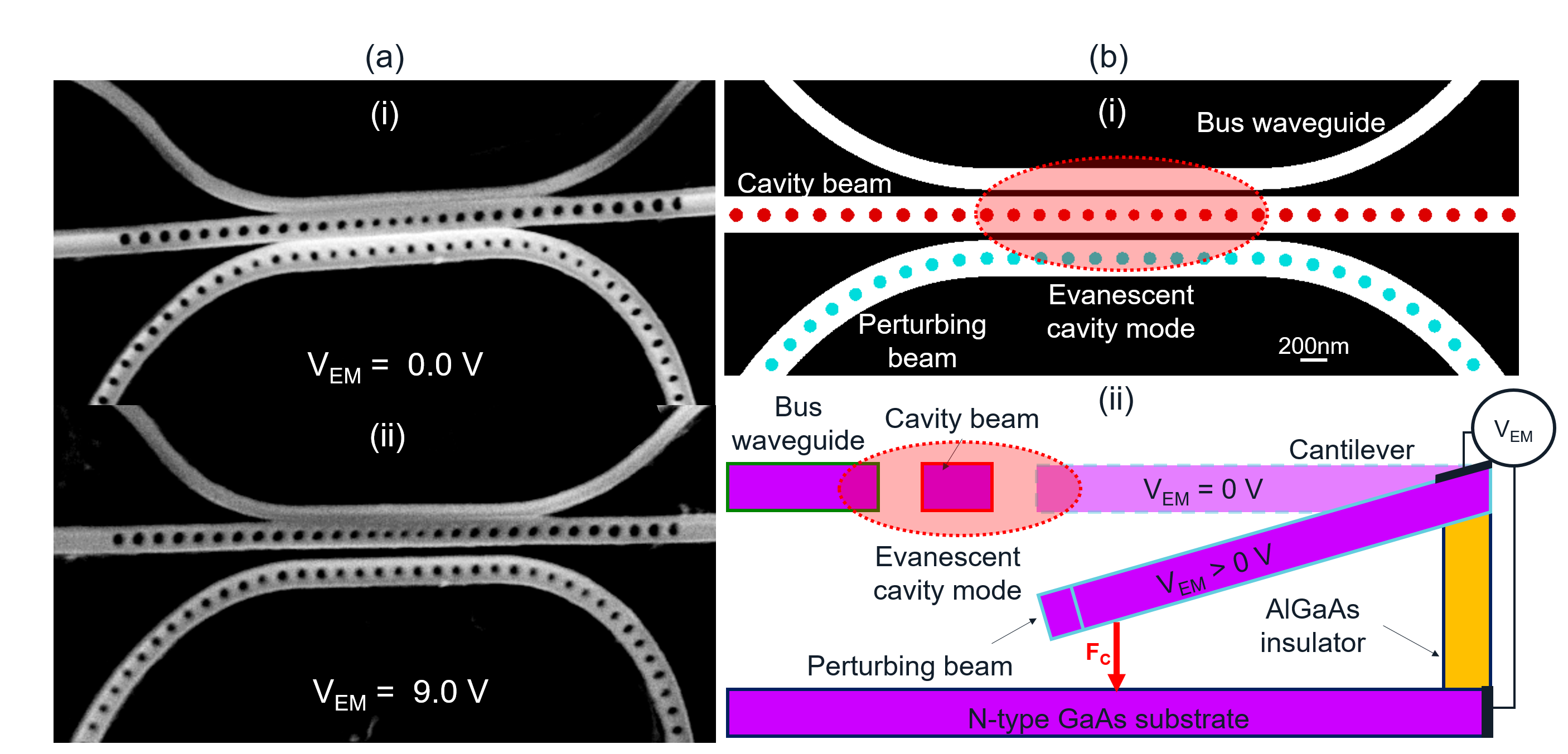}
\caption{(a) SEM images of cavity/bus waveguide region at (i) $V_{EM} = 0.0$ V and (ii) $9.0$ V showing the downward displacement of the perturbing beam at 9.0 V due to the electrostatic actuation of the cantilever. (b) (i) Illustration of the spatial overlap between the evanescent cavity mode and the perturbing beam which is actuated to tune the cavity wavelength. (ii) Schematic of device operation depicting the application of bias between the cantilever and the substrate creating an electro-static force which deflects the cantilever towards the substrate.}
\label{fig:2}

\end{figure}\autoref{fig:2} illustrates the operating principle of the device. When $V_{EM}$ is applied, an electric field is produced between the cantilever and the substrate. This effectively creates a capacitor with charges accumulating on each of the "plates". As $V_{EM}$ is increased, the electric field strength increases due to the build up of oppositely signed charges on each plate. At a critical charge accumulation, such that the force of electrostatic attraction is greater than the restoring force produced by the elastic strain of the cantilever, the suspended structure will begin to deflect towards the substrate. As the cantilever actuates, the perturbing beam gradually shifts out of the evanescent cavity mode. With increasing deflection, the influence of the perturbing beam on the cavity mode index diminishes, thus, reducing the effective mode index resulting in a  blue-shift of the cavity mode.

The displacement of the cantilever is governed by the following relationship~\cite{cantspringconstantderevation}:\\
\begin{equation}
\left( d_0 - d \right) = \frac{2 \epsilon L^4}{E d^2 t^3}V^2 \label{equ:displacementequationfull}
\end{equation}Where $V$ is the applied bias, $d_0$ is the distance between the cantilever and the substrate at equilibrium ($V = 0$), $d$ is the displacement, $\epsilon$ is the permittivity of the material between the substrate and the membrane, $L$ is the length of the cantilever and $E$ is the Young's modulus of the cantilever material. The maximum displacement of the perturbing beam from the equilibrium point is therefore reached at the pull-in voltage $V_{pi}$, the point where the electrostatic and restoring forces are balanced. This corresponds to a displacement of $d_{max} = \frac{1}{3}T_{SL}$ where $T_{SL}$ is the thickness of the sacrificial layer. Above $V_{pi}$, the electrostatic force overcomes the restoring force, causing the cantilever to collapse and adhere to the substrate. If $V_{EM}$ is reduced to zero, the cantilever can recover to its initial state if the restoring force is greater than the Van der Waals force between the substrate and the cantilever. 
Using \autoref{equ:displacementequationfull} we calculate the theoretical pull-in voltage of our devices to be $V_{pi} = 2.50$ V. \autoref{fig:2} (a) and (b) shows angled SEM images of a device at $V_{EM} = $ 0.0 V and 9.0 V ,respectively, where the displacement of the perturbing beam relative to the cavity beam is evident. Notably, the beam does not collapse at 9.0 V despite this voltage exceeding the theoretical pull-in voltage, which we attribute to resistance in the circuit used to bias the cantilever. The actuation voltage is reduced when operating at cryogenic temperatures due to the reduced resistance. At cryogenic temperatures, the pull-in voltage becomes $V_{pi} \sim 6.5$ V.
 
\subsection{Cavity Tuning Characteristics}
\label{sec:2.2}
\begin{figure}[h!]
\centering
\includegraphics[width=1\textwidth]{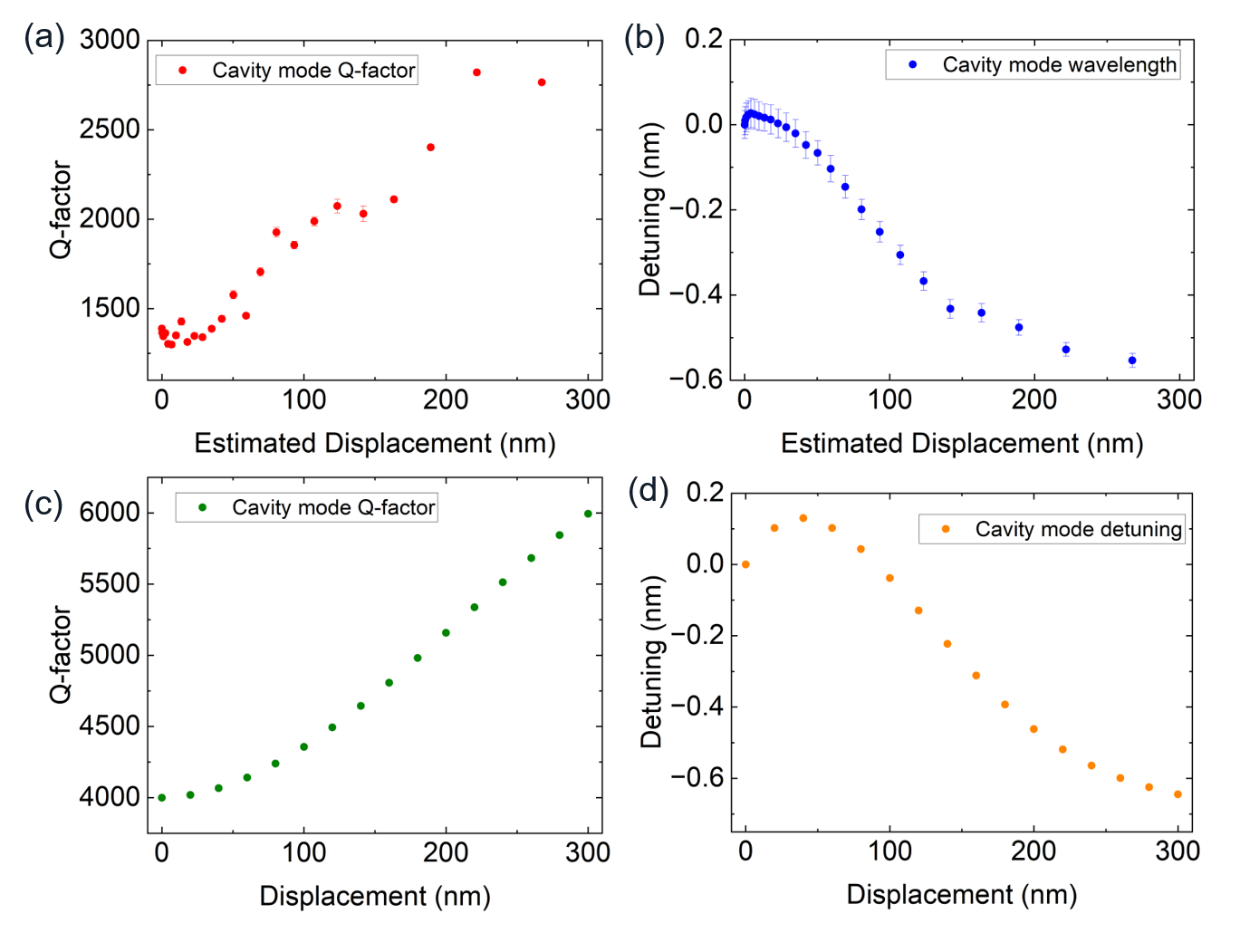}
\caption{Device 1. (a,c) Experimental and simulated relationship between Q-factor and perturbing beam displacement, respectively (b,d) Experimental and simulated relationship between cavity mode detuning relative to initial cavity wavelength at $V=0$, respectively.}
\label{fig:3}
\end{figure}
To study the tuning behaviour of the structures, the wavelength and Q-factor of the cavity mode were measured as a function of $V_{EM}$ using a micro-photo-luminescence set-up. An above-band laser (808 nm) was used to excite the broadband QD ensemble in the cavity region, thereby illuminating the cavity mode. The displacement of the perturbing beam was estimated using \autoref{equ:displacementequationfull}, with a small empirical offset in voltage of $V_{off} = 3.8$ V to account for the circuit resistance. 
\autoref{fig:3} shows a comparison between the experimentally measured and simulated tuning behaviour of a typical cavity (Device 1). Overall, a good agreement between the simulated and experimentally measured structures can be seen for both the Q-factor (\autoref{fig:3} (a) and (c)) and wavelength dependence (\autoref{fig:3} (b) and (d)) as a function of displacement. To reproduce the lower Q-factors seen experimentally compared to the optimised simulated cavities - which exhibit Q-factors in excess of $1\times 10^6$ - we introduce large scattering losses from the cavity and perturbing beams by modulating the size and positions of the air holes within the beams. This reduces the Q-factor of the simulated structures to the same order of magnitude as the measured devices. The Q-factor increases slowly at displacements between 0 and 50 nm, before increasing more rapidly at larger displacements above 50 nm. The relationship between the Q-factor and the displacement becomes roughly linear in the simulation at displacements larger than 150 nm. The overall increase in the Q-factor is smaller in the experimental data compared to the simulated device. This may be due to the lower starting Q-factor in the experimental device. As the Q-factor continues to increase right up to the end of the range, this implies that scattering losses from the perturbing beam are the dominant loss mechanism (see \autoref{sec:losses} for more detail). These are likely due to sub-optimal air hole radii in the perturbing beam and sidewall roughness in the fabricated device. In \autoref{fig:3} (b) a small red-shift in the cavity wavelength is observed at small perturbing beam displacements. This behaviour is also seen in the simulations, although the magnitude of the redshift is larger in the simulated device. This shift may be due to the initial displacement shifting the perturbing beam through an anti-node of the cavity mode field, thus increasing the effective index of the mode producing a red-shift in the cavity mode wavelength. The maximum observed tuning range of this device was $\Delta \lambda = 0.55$ nm. This range corresponds to a separation of approximately 85 nm between the cavity and perturbing beams in simulation. This agrees with the typical separation extracted from SEM images of the fabricated devices (see \autoref{sec:1DPhCCtuning} for further details).
\begin{figure}[t!]
\centering
\includegraphics[width=1\textwidth]{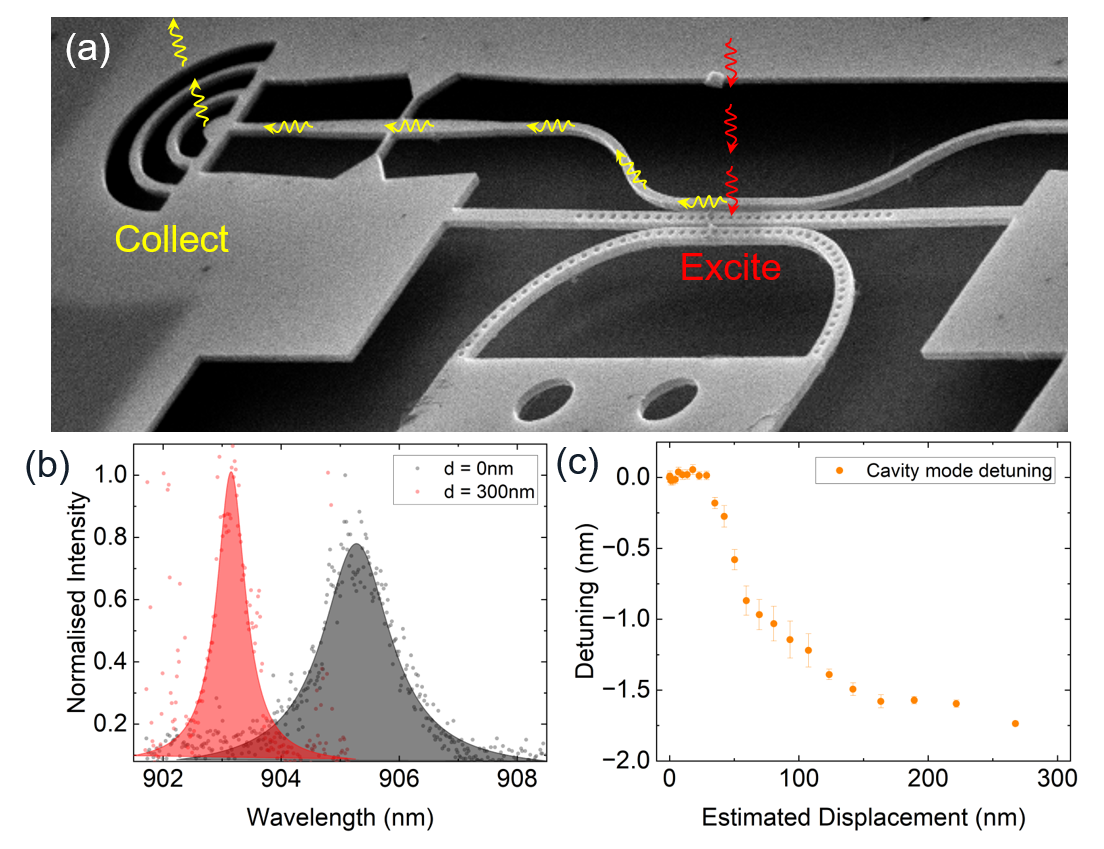}
\caption{Device 2. (a) Annotated SEM image detailing the excitation and collection scheme of the measurements in (b) and (c). (b) Cavity mode PL spectra of device 2 with perturbing beam displacements of 0 nm (grey) and 300 nm (red). (c) Experimentally measured relationship between the cavity mode detuning and the perturbing beam displacement for Device 2 demonstrating a maximum tuning range of 1.8 nm.}
\label{fig:4}
\end{figure}
\\To demonstrate bus waveguide coupling, the cavity detuning characteristics of a second cavity (Device 2) were measured by exciting the cavity mode with an above-band laser and collecting the light scattered out of the structure by the output coupler which terminates the bus waveguide as shown in \autoref{fig:4} (a).
\autoref{fig:4} (b) shows the PL spectra of the cavity mode with the cantilever displaced by 0 nm (black) and 300 nm (red), using the excitation and collection scheme depicted in \autoref{fig:4} (a). Device 2 exhibits the same blue-shifting behaviour as observed for Device 1. However, the maximum detuning observed from this device was larger at $\Delta \lambda = 1.8$ nm with the detuning vs displacement dependence being shown in \autoref{fig:4} (c). This tuning range corresponds to a separation of approximately 57.5 nm between the perturbing and cavity beams which is estimated from the simulated relationship between detuning range and in-plane separation between the cavity and perturbing beams (see \autoref{sec:1DPhCCtuning}). The difference in separation between Device 1 and Device 2 is a result of slight variations in the design parameters. This large difference in tuning range (0.55 nm vs 1.8 nm) demonstrates the sensitivity of the device to the separation between the beams (85 nm vs 57.5 nm). We also observe the expected increase in the Q-factor of the cavity mode as the cantilever is deflected as seen in the first device, increasing from $Q_0 = 640$ at a displacement of 0 nm, to $Q_{300} = 1530$ at a displacement of 300 nm similar to the behaviour observed in \autoref{fig:3} (a) and (c).
The results from \autoref{fig:3} and \autoref{fig:4} clearly demonstrate the robust, voltage-controllable cavity mode wavelength tuning capabilities of our device and how the light can be extracted off-chip via a side-coupled bus waveguide terminating in a grating output coupler. Furthermore, the indirect cavity tuning does not significantly degrade the Q-factor of the cavity. Our devices therefore exhibit key attributes necessary for the effective scale-up of QD-cavity based systems.
\subsection{Quantum dot tuning into resonance with cavity}
\begin{figure}[h!]
\centering
\includegraphics[width=1\textwidth]{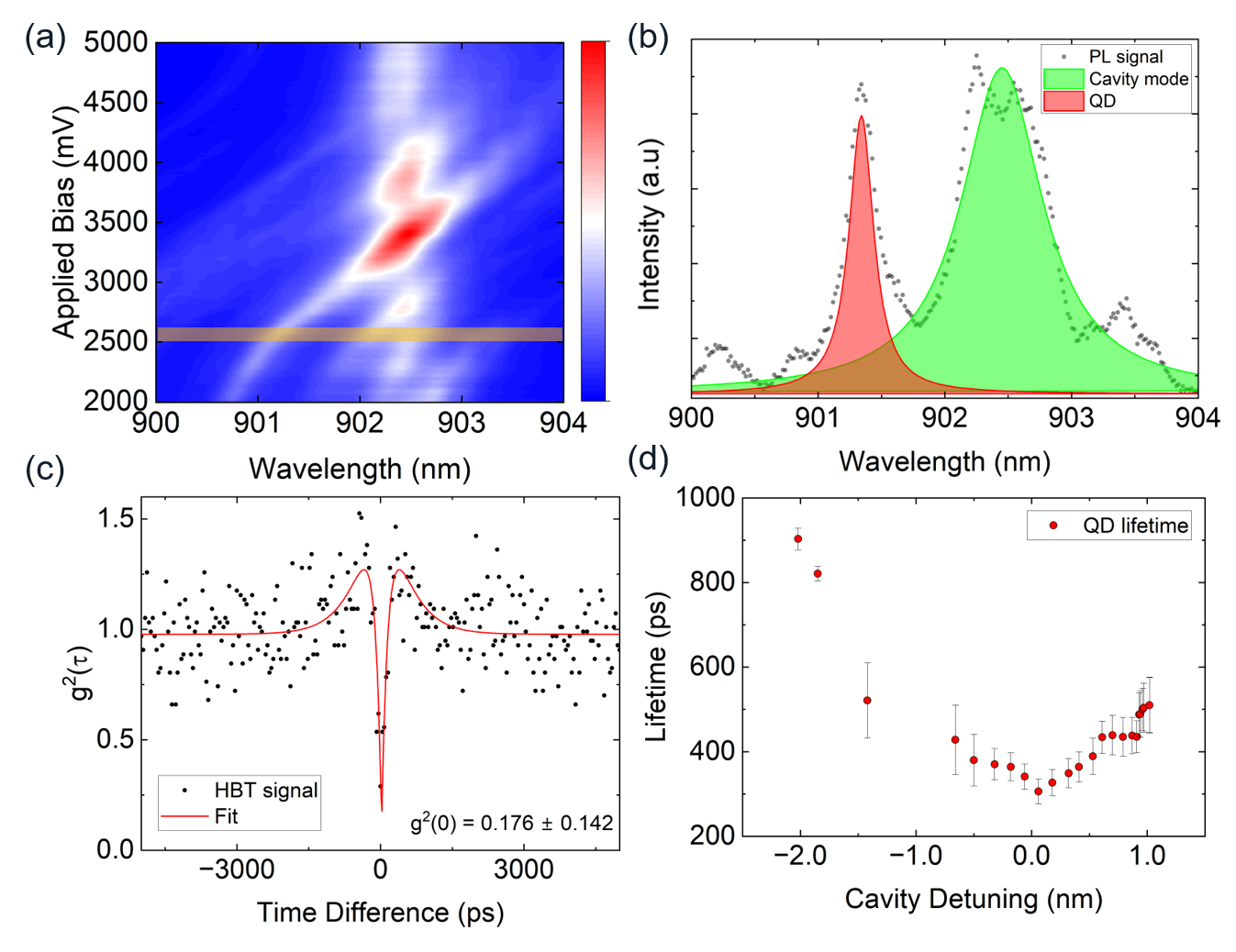}
\caption{Device 3. (a) QD emission line electrically tuned via applied bias through a cavity mode $\sim 902.5$ nm. (b) Single QD PL-spectrum from (a) (shaded orange region). (c) Second-order correlation measurement of the emission from the QD in (b). (d) Non-resonant PL lifetime of the QD as a function of cavity detuning.}
\label{fig:5}
\end{figure}
To demonstrate further the potential of our device for quantum technologies applications, we explored the behaviour of a single QD located in a third cavity (Device 3) from the same wafer. \autoref{fig:5} (a) shows the PL spectrum of a QD state tuning as a function of the applied bias when exciting and collecting directly above the cavity. A clear enhancement in emission intensity is observed when the QD becomes degenerate with a weakly excited cavity mode at $\sim 902.5$ nm, a signature of Purcell enhancement.
Furthermore, a single PL spectrum corresponding to the orange shaded region in \autoref{fig:5} (a) is plotted in \autoref{fig:5} (b). This shows the QD photo-luminescence signal (red curve) with the cavity mode (green curve), as distinct spectral features. By filtering for the QD emission only and removing the influence of the near-resonant cavity mode, we investigated the photon statistics of the QD, represented by $g^{(2)}(\tau)$. As shown in \autoref{fig:5} (c), a clear anti-bunching dip at zero delay time ($\tau=0$) is observed, with $ g^{(2)}(0) = 0.176 \pm 0.142 $, well below the threshold of 50\% for single-photon emission. 
\autoref{fig:5} (d) further explores the interaction between the QD and the cavity by plotting the non-resonant PL lifetime of the QD emission as a function of detuning from the cavity mode. A substantial reduction in radiative lifetime is observed when the QD is in resonance with the cavity, reaching a minimum of approximately 350 ps at zero detuning (limited by APD response time).

\section{Conclusion}
This work demonstrates the successful integration of quantum dots (QDs) with tunable photonic-crystal cavities on a GaAs platform, enabling precise control over QD-cavity coupling. By combining voltage-controlled QD tuning with independent cavity mode adjustment via electro-mechanically actuated cantilevers, we achieve robust and repeatable resonance alignment between QD emission and single cavity modes. This independent tuning mitigates the challenges posed by the stochastic nature of QD positioning and fabrication-induced variations in cavity dimensions—a key barrier to scalable QD-cavity coupled systems.

The good agreement between simulated and experimentally measured tunable cavity Q-factors and wavelengths provides a reliable framework for realising this design as a scalable platform for multiple cavity-coupled systems. Furthermore, the device design, employing a 1D-PhCC with a low mode volume, ensures efficient coupling between QDs and cavity modes while minimising losses. The implementation of cantilever-based tuning offers a localised method for cavity wavelength adjustment, preserving the intrinsic optical properties of the QDs — a challenge in many other systems.

The experimental results highlight several key achievements. First, a large, voltage controllable tuning cavity mode tuning range is observed in a side-coupled geometry. This paves the way for scale-up to multiple waveguide-mediated coupled-cavity devices, with efficient on and off-chip coupling of photons. Second, Purcell enhancement is observed through increased PL intensity and reduced PL lifetime when QDs are resonant with cavity modes. Additionally, single-photon emission with marked anti-bunching characteristics is demonstrated, with photon purity limited only by material losses and non-resonant excitation conditions. Finally, we achieve such strong single-photon behaviour within a tunable cantilever system, enabling more intricate studies of QD-cavity interactions in future work.

Overall, this work establishes a robust framework for achieving precise QD-cavity coupling in semiconductor systems, paving the way for advancements in quantum light sources and cavity-enhanced photon emission. These results lay the foundation for future exploration of complex quantum systems, including multi-QD cavity-coupled systems for the realisation of cavity-based optical switches in photonic circuit designs and the investigation of QD - QD interactions within coupled-cavity environments.

\section{Methods}
The wafer was grown on a (100) GaAs substrate via molecular beam epitaxy. The top 300nm of the substrate is n-doped with silicon atoms at a density of $2\times 10^{18}$ cm$^{-2}$. A 1.15 $\mu$m layer of Al$_{0.6}$Ga$_{0.4}$As is grown on top of the substrate, the top 200 nm of which is n-doped with silicon atoms at a density of $2\times 10^{18}$ cm$^{-2}$. This sacrificial layer is selectively etched away in the device fabrication process. Next, the device membrane containing embedded InAs QDs is grown. The primary material of the 170 nm thick layer is GaAs, in addition, two Al$_{0.3}$Ga$_{0.7}$As quantum well barrier layers are included either side of the QD layer to increase the electrical tuning range of the QD emission energies. The barrier above/below the QD layer is 30/50 nm thick, respectively. The QDs are grown by the Stranski–Krastanov method and are located at the centre of the membrane. The top/bottom 30 nm of the membrane are p/n-doped with carbon/silicon atoms at a density of $2\times 10^{19}$ cm$^{-2}$ and $2\times 10^{18}$ cm$^{-2}$, respectively.\\The nanostructures were patterned into a SiO$_{2}$ resist via electron beam lithography and were subsequently transferred into the membrane through an inductively-coupled plasma etch. The wafer was then exposed to HF acid to selectively etch away the sacrificial layer and produce free-standing structures. Finally, the wafer underwent a critical-point dry process to preserve the structures during the removal of the residual acid. To create the diode structures as depicted in \autoref{fig:1} (c) mesas of the were etched into the wafer at a depth of 150 nm for the QD tuning and 1300 nm for the cantilever actuation. The p and n-type layers were then patterned with Ti/Au contacts which were connected to individual pins on the chip carrier to enable separate bias' to be applied to the QD and cantilever layers.\\The sample was placed in a low pressure helium exchange-gas bath cryostat on 3-axis piezo-electric stages. A pair of achromatic doublets and an aspheric lens are used to efficiently focus light onto and collect emission from the sample.\\The optical measurements were conducted using a standard confocal micro-photo-luminescence set-up where the QDs and cavity modes were excited non-resonantly with an above band (808 nm) continuous-wave laser. The spectra were recorded using a silicon charge-coupled detector positioned after a grating spectrometer. The time resolved data shown in \autoref{fig:5} (d) were obtained via excitation from a pulsed Ti:S laser operating above band ($\sim$810 nm) at $\sim$80 MHz repetition rate. The signal was detected using an APD with a timing resolution of 350 ps and dead time of 22 ns connected to a time-tagger with a bin width of 50 ps.

\section{Author Information}
\textbf{Corresponding author}
\begin{itemize}
    \item Luke Brunswick - \textit{School of Mathematical and Physical Sciences, University of Sheffield, UK}; https://orcid.org/0000-0001-8157-7255; l.brunswick@sheffield.ac.uk
\end{itemize}
\textbf{Authors}
\begin{itemize}
    \item Luke Hallacy - \textit{School of Mathematical and Physical Sciences, University of Sheffield, UK}
    \item Ren\'{e} Dost - \textit{School of Mathematical and Physical Sciences, University of Sheffield, UK}
    \item Edmund Clarke - \textit{School of Electronic and Electrical Engineering, University of Sheffield, UK}
    \item Maurice Skolnick - \textit{School of Mathematical and Physical Sciences, University of Sheffield, UK}
    \item Luke Wilson - \textit{School of Mathematical and Physical Sciences, University of Sheffield, UK}
\end{itemize}\textbf{Funding}\\This work was funded by the Engineering and Physical Sciences Research Council (EP/V026496/1).\\ \textbf{Notes}\\
The authors declare no competing financial interests.
\begin{acknowledgement}

The authors thank Andrew Foster for helpful discussions.

\end{acknowledgement}
\begin{suppinfo}

The following files are available free of charge.
\begin{itemize}
  \item Supporting information: Details of optimisation of cavity design, tolerance of cavity design to fabrication disorder, simulated tuning range of device, effect of material loss on cavity performance, side-coupling efficiency, schematic of the set-up and details of the cross-correlation measurement.
\end{itemize}

\end{suppinfo}
\newpage
\bibliography{Master}


\clearpage

\begin{center}
\textbf{\large Supporting Information: Electro-mechanically tunable, waveguide-coupled photonic-crystal cavities with embedded quantum dots}

\end{center}
\newpage
\setcounter{equation}{0}
\setcounter{figure}{0}
\setcounter{table}{0}
\setcounter{page}{1}
\setcounter{section}{0}
\makeatletter
\renewcommand{\theequation}{S\arabic{equation}}
\renewcommand{\thefigure}{S\arabic{figure}}

\renewcommand{\thesection}{S\arabic{section}}

\newpage

\section{Nanobeam cavity optimisation \label{SI:cavity_design}} 

This section will contain details of the simulation study performed to optimise the design of the 1D photonic-crystal cavity (PhCC) used in the main text. All data is reproduced from~\cite{LBthesis}.

\subsection{Uncoupled, lossless cavity optimisation}
The design of the 1D-PhCC was optimised through a set of parameter sweeps conducted in Lumerical 3D-FDTD simulations. Each of the parameters from \autoref{fig:1} (d) in the main text, were swept over a large range while the other parameters were held constant. 

\begin{figure}[t!]
    \centering
    \includegraphics[width=1\textwidth]{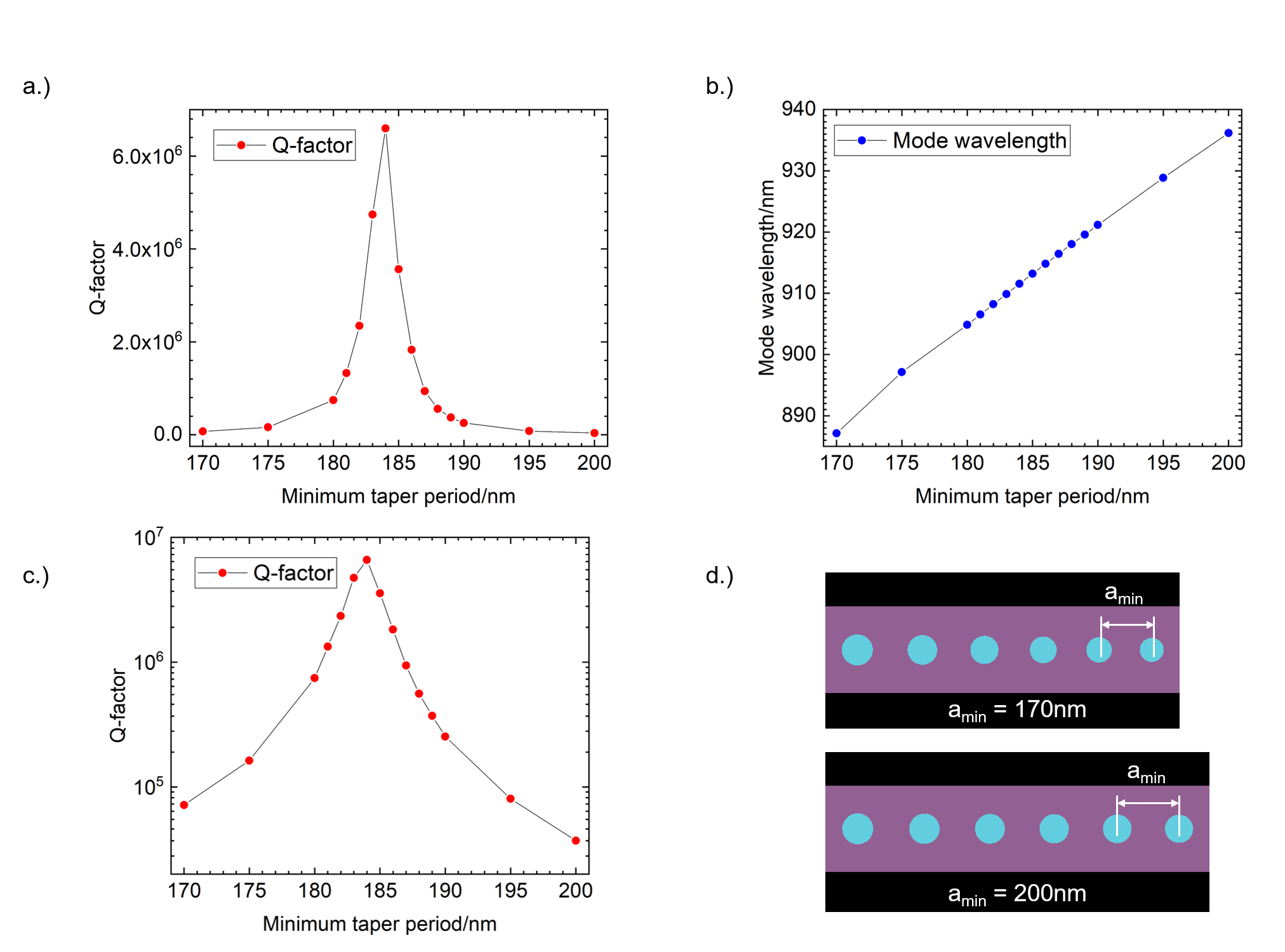}
    \caption{Results from the parameter sweep simulations of $a_{min}$ for (a) Q-factor and (b) cavity mode wavelength. (c) Q-factor result presented on a logarithmic scale. (d) Visualisation of the effect on the structure of changing $a_{min}$.}
    \label{fig:amin}
\end{figure}

 \begin{figure}[t!]
    \centering
    \includegraphics[width=1\textwidth]{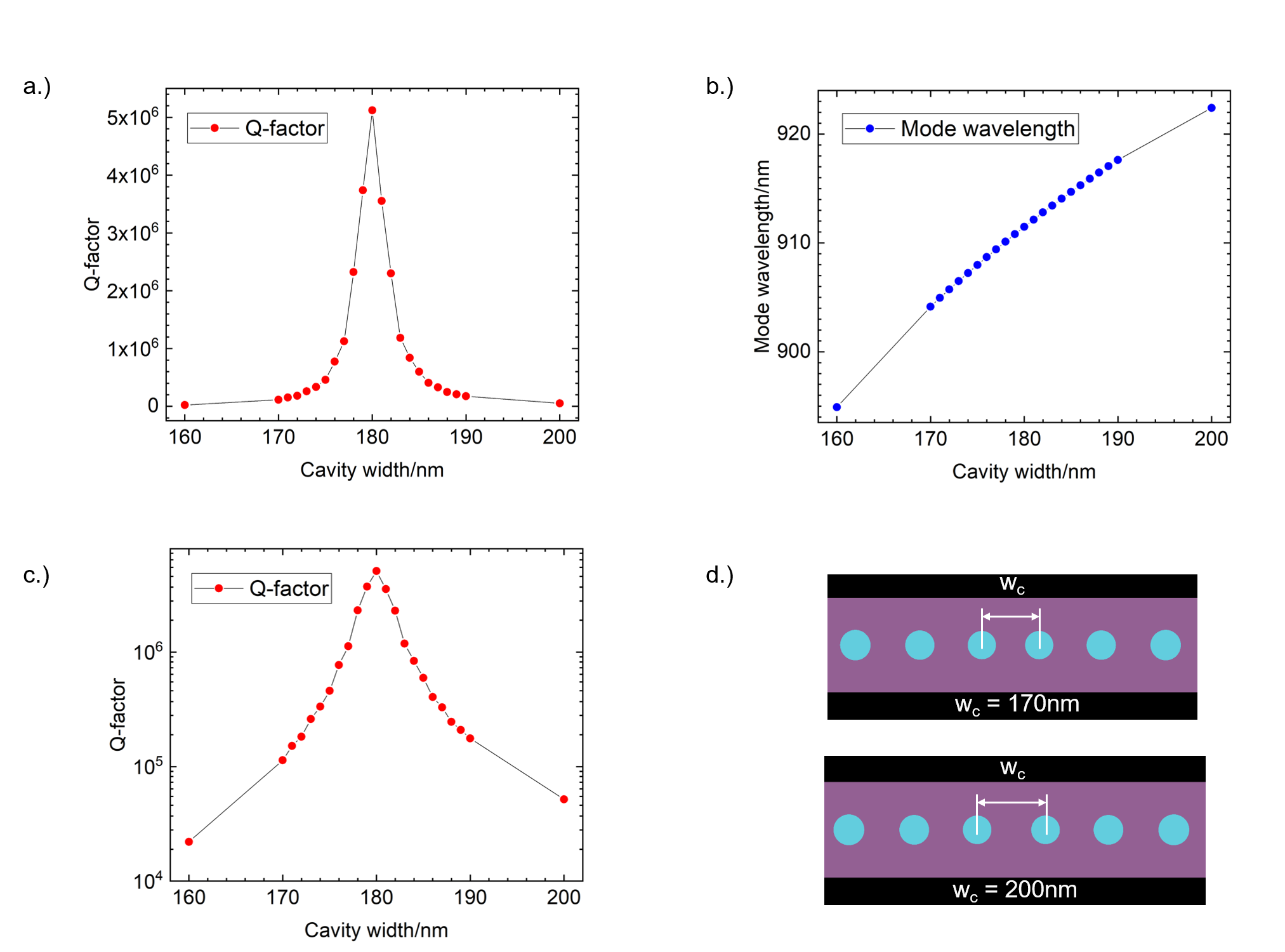}
    \caption{Results from the parameter sweep simulations of $c_{w}$ for (a) Q-factor and (b) cavity mode wavelength. (c) Q-factor result presented on a logarithmic scale. (d) Visualisation of the effect on the structure of changing $c_{w}$.}
    \label{fig:cw}
\end{figure}

The two most sensitive parameters in the design were the minimum period of the photonic-crystal ($a_{min}$) and the cavity width ($c_w$). The results from the parameter sweeps of $a_{min}$ and $c_w$ are shown in \autoref{fig:amin} and \autoref{fig:cw}, respectively. Both parameters exhibit a strong resonance-like behaviour about the optimal value in their Q-factor dependence, demonstrating the sensitivity of these parameters. This is especially true of $a_{min}$, as fabrication imperfections largely manifest in air hole radii fluctuations, rather than positional errors.\\

\begin{figure}[t!]
    \centering
    \includegraphics[width=1\textwidth]{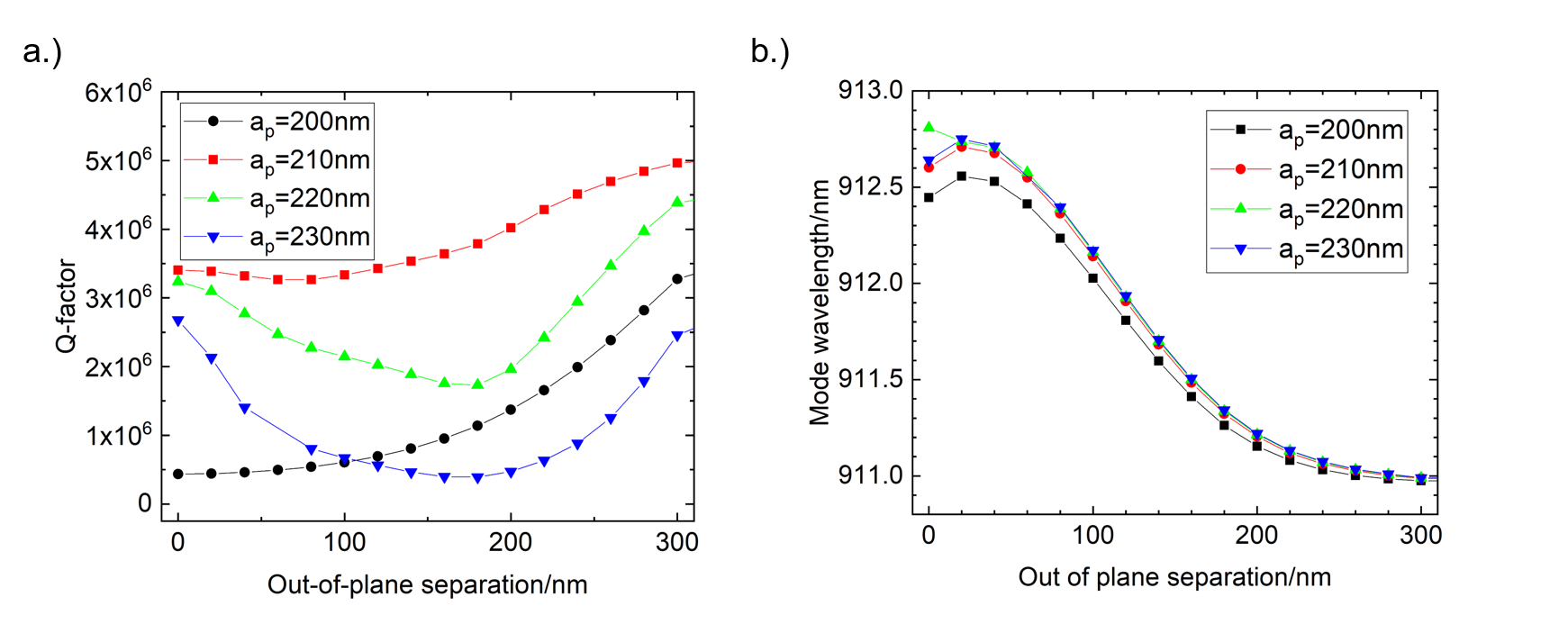}
    \caption{Results from the parameter sweep simulations of $a_{p}$ for (a) Q-factor and (b) cavity mode wavelength dependence on out-of-plane separation between the cavity and perturbing beam.}
    \label{fig:ap}
\end{figure}

Further to this, the parameters of the photonic-crystal in the perturbing beam were optimised to reduce losses associated with the presence of the perturbing beam in the evanescent cavity field. From \autoref{fig:ap}, we can see that the reduction in Q-factor caused by the actuation of the perturbing beam can be eliminated by using the optimal photonic-crystal period. A value of $a_{p} = 210$ nm was found to be optimal in our simulations. This is slightly smaller than the value of $a = 220$ nm used for the cavity beams. This reduction in the period does not significantly alter the achievable tuning range of the device as shown by \autoref{fig:ap} (b).

\newpage
\subsection{Nanobeam cavity Fabrication Tolerance}
\begin{figure}[h!]
    \centering
    \includegraphics[width=1\textwidth]{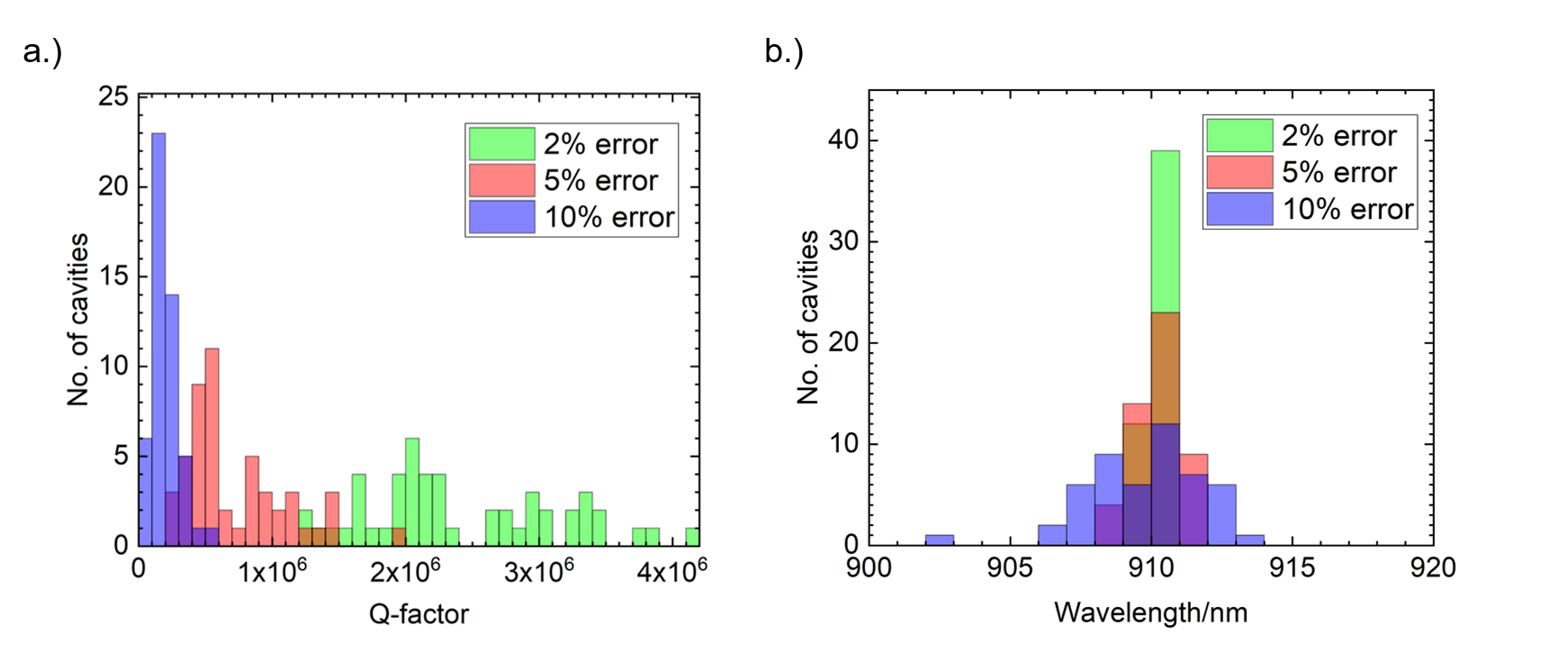}
    \caption{Results from the Monte Carlo simulations for 2\% (green), 5\% (red) and 10\% (blue) error on the air hole radii.}
    \label{fig:MC}
\end{figure}

To assess how robust to fabrication imperfections the optimised cavity design was, a Monte Carlo simulation study was conducted for three different air hole radii errors: 2\%, 5\% and 10\%. In each case, 50 different devices were simulated where the radius of each air hole was modified randomly within a uniform distribution. \autoref{fig:MC} (a) and (b) show the effect of the three different levels of fabrication disorder on the Q-factor and cavity mode wavelength, respectively. 
\begin{table}[h!]
\centering
\begin{tabular}{|l|c|c|c|}
\hline
\textbf{Error} & \textbf{$Q_{Min}$} & \textbf{$Q_{Med}$} & \textbf{$\sigma _{\lambda}$}  \\
\hline
2\% & $1.26\times 10^{6}$  &  $2.20\times 10^{6}$ & $0.29$ nm\\
5\% & $2.69\times 10^{5}$ & $5.56\times 10^{5}$ & 0.74 nm\\
10\% & $7.47\times 10^{4}$ & $1.82\times 10^{5}$ & 1.99 nm\\
\hline
\end{tabular}
\caption{Qualitative results from \autoref{fig:MC}}
\label{tab:MC_SD}
\end{table}
\autoref{tab:MC_SD} shows the numerical results from the Monte Carlo simulations. As the level of error increases, the minimum and median Q-factor values reduce, while the standard deviation in the cavity mode wavelength distribution increases.\\

\begin{figure}[t!]
    \centering
    \includegraphics[width=1\textwidth]{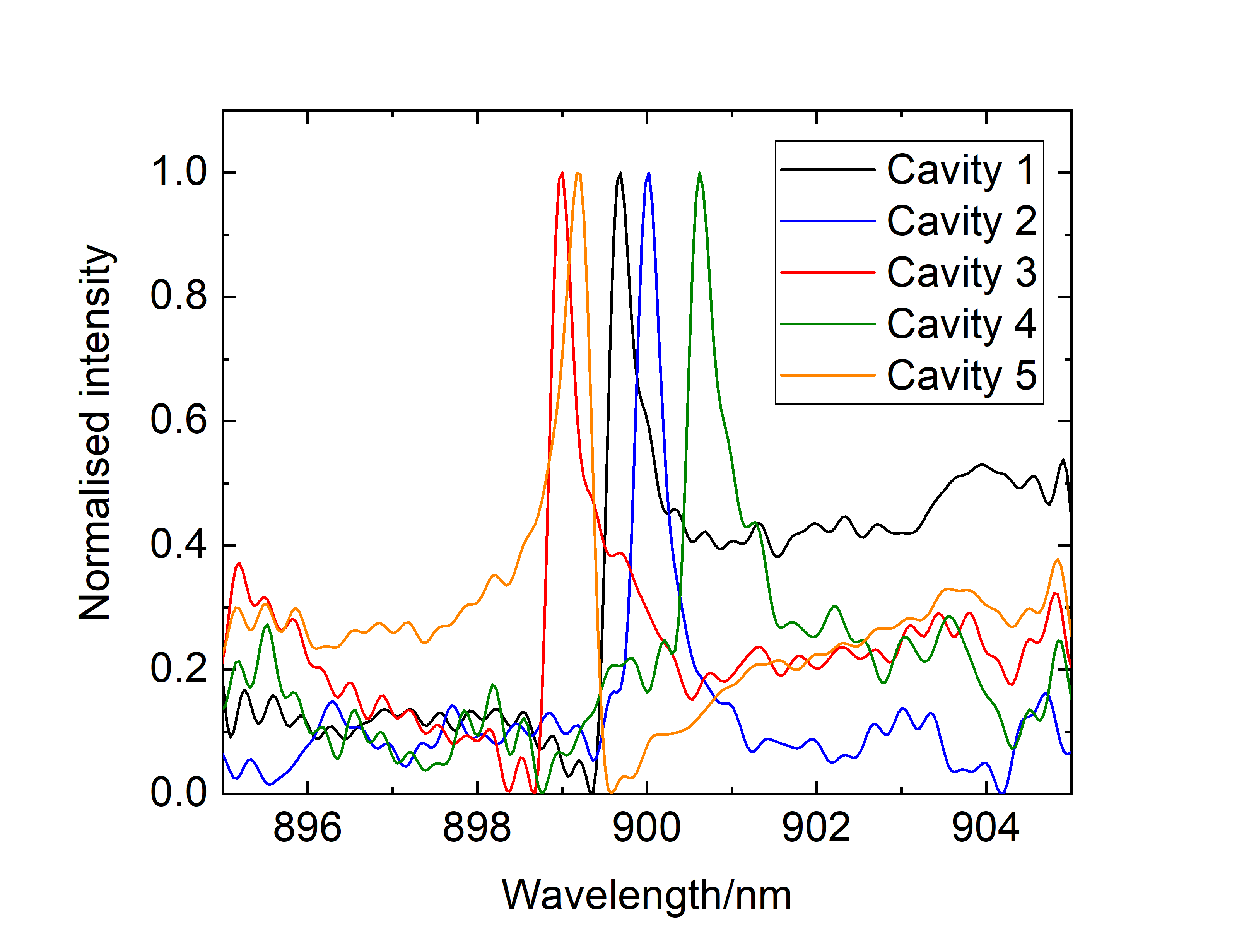}
    \caption{Overlapped white light reflection spectra from the set of like cavities with the lowest standard deviation in cavity mode wavelength distribution.}
    \label{fig:fabSD}
\end{figure} Prior to fabricating a sample combining the cavity, perturbing beam and waveguide elements into a single device, a sample containing just cavity devices was made. A range of different cavity parameter sets were present on the sample to discern which parameter set gave the best cavity performance in experiment. \autoref{fig:fabSD} shows the distribution of cavity mode wavelengths from the parameter set which exhibited the lowest standard deviation of $\sigma _{\lambda} = 0.6$ nm, equating to an error of 2-5\% on the air hole radius.
\newpage
\section{1D-PhCC cavity tuning}
\label{sec:1DPhCCtuning}
\subsection{Electrical isolation}
Unfortunately, due to electrical cross-talk between the diodes, we were unable to tune both the cavity and QD emission simultaneously in a controllable manner. We attribute this cross-talk to the insufficient depth of the etched isolation trenches in the device membrane. The cross-talk could be eliminated by increasing the depth of the isolation trenches to fully etch through the membrane, rather than just the p-type layer.
\subsection{Tuning range}
\label{sec:tuningrange}
\begin{figure}[t!]
\centering
\includegraphics[width=1\textwidth]{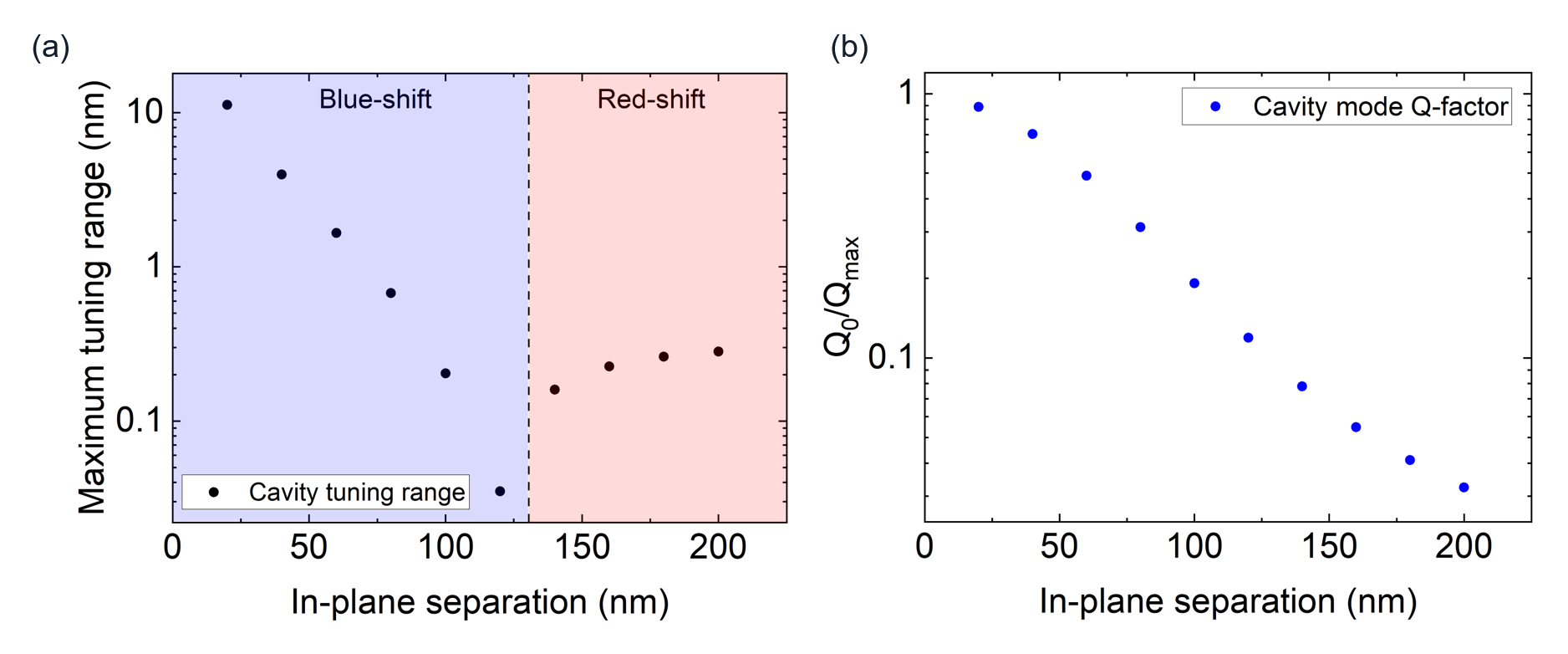}
\caption{(a) Simulated relationship between the separation between the cavity and perturbing beam and the maximum tuning range of the cavity mode wavelength. (b) Simulated relationship between the Q-factor dip and the separation between the cavity and perturbing beam}
\label{fig:ips}
\end{figure}The maximum achievable tuning range of our device is strongly dependent on the separation between the cavity and perturbing beams. To quantify this, a simulation study was conducted, the results of which are presented in \autoref{fig:ips}. The maximum tuning range of the device falls off exponentially as the separation between the cavity and perturbing beam increases up until a separation of $\sim 130$ nm. After this, the cavity mode wavelength no longer blue-shifts as the cantilever is deflected and instead, red-shifts a small amount. The Q-factor dip decreases in a roughly exponential manner as the separation increases. 
\begin{figure}
    \centering
    \includegraphics[width=0.5\linewidth]{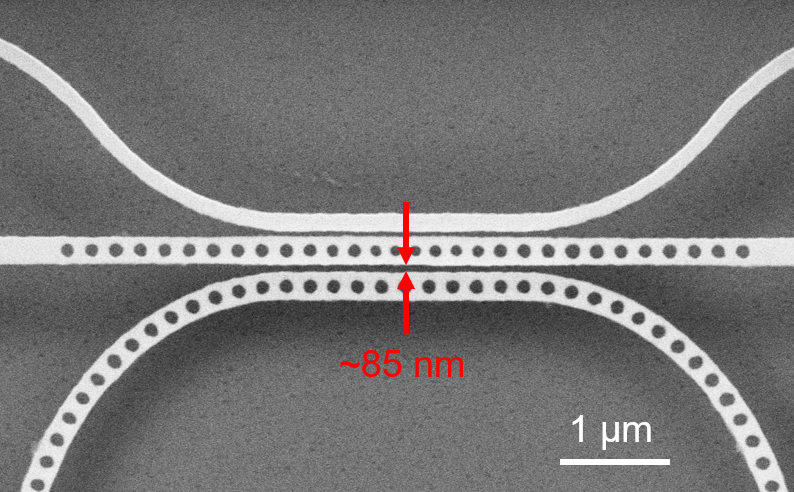}
    \caption{SEM image showing a typical cavity/waveguide/perturbing beam interface with the separation between the perturbing and cavity beam measured to be $\sim 85 $nm.}
    \label{fig:measuredSEM}
\end{figure}\autoref{fig:measuredSEM} shows an SEM image of a typical structure from the measured sample. The measured separation between the cavity and perturbing beams is $\sim 85$ nm, which corresponds to a simulated tuning range of 0.55 nm. This result is consistent with the results in \autoref{fig:4}.
\subsection{Effect of material loss on Q-factor}
\label{sec:losses}
To study the effect of material absorption loss on the Q-factor of the cavity, the simulations used to produce \autoref{fig:3} (c) \& (d) in the main text were repeated with the imaginary refractive index of the material set to $n_i = 1\times 10^{-4}$ and $n_i = 5\times 10^{-4}$. 
\begin{figure}
    \centering
    \includegraphics[width=1\linewidth]{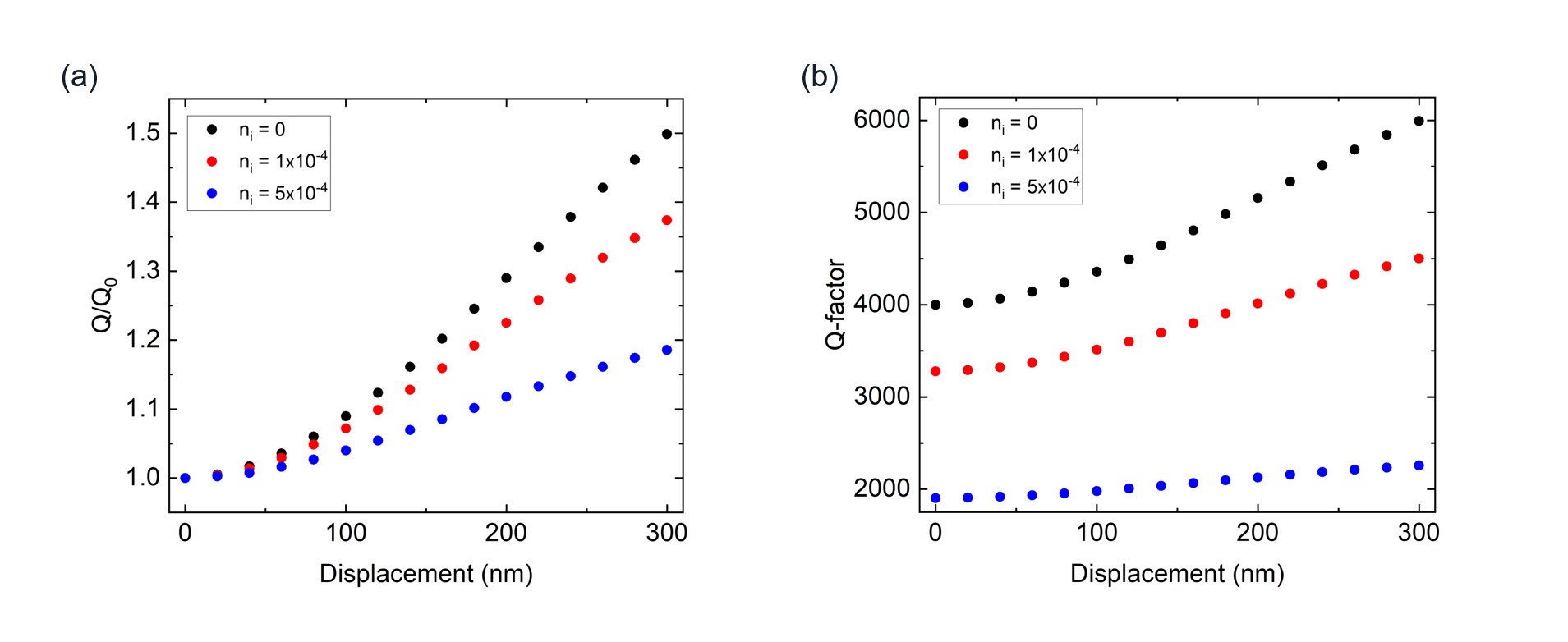}
    \caption{Relationship between the perturbing beam displacement and the (a) relative, (b) absolute, cavity Q-factor for an imaginary refractive index of $n_i = 0$ (black circles), $n_i = 1\times10^{-4}$ (red circles) and $n_i = 5\times10^{-4}$ (blue circles).}
    \label{fig:materialloss}
\end{figure}
\autoref{fig:materialloss} (a) and (b) show the relationship between the perturbing beam displacement and the relative and absolute cavity Q-factor, respectively. As the value of $n_i$ increases, both the starting Q-factor ($Q_0$) and the relative increase in Q-factor decrease. The Q-factor trend also changes, starting to plateau more noticeably at higher values of $n_i$. We can infer from this, that the main source of loss in our fabricated device is scattering loss due to the presence of the perturbing beam in the evanescent cavity field.   
\newpage
\section{Bus waveguide coupling}
\label{sec:buswgcoupling}
The devices simulated in this section were based on an optimised, lossless cavity with a starting $Q \sim5\times10^6$. 
\begin{figure}
    \centering
    \includegraphics[width=1\linewidth]{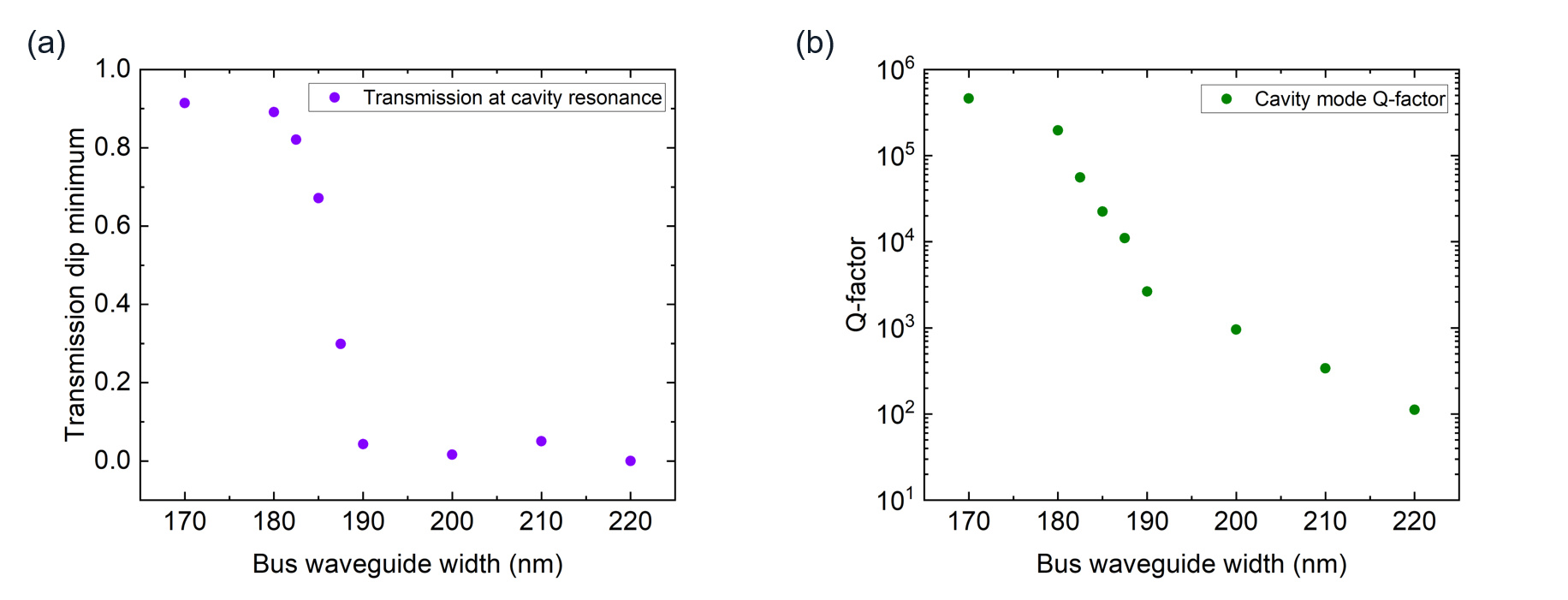}
    \caption{Simulations showing the relationship between the width of the bus waveguide and (a) end-to-end transmission through the waveguide (b) cavity mode Q-factor for a fixed coupling gap of $d_0 = 60$ nm.}
    \label{fig:bwgcoupling}
\end{figure}
\autoref{fig:bwgcoupling} (a) and (b) show the relationship between the width of the bus waveguide and the transmission through the device and cavity mode Q-factor for a fixed coupling gap, respectively. The magnitude of the transmission dip increases dramatically in between widths of 180 and 190 nm. This is due to the increased k-space overlap of the cavity and waveguide modes increasing the coupling efficiency between the two modes. This increase in coupling efficiency is also reflected in the reduction of the cavity Q-factor by 3 orders of magnitude over the same range.
\begin{figure}
    \centering
    \includegraphics[width=1\linewidth]{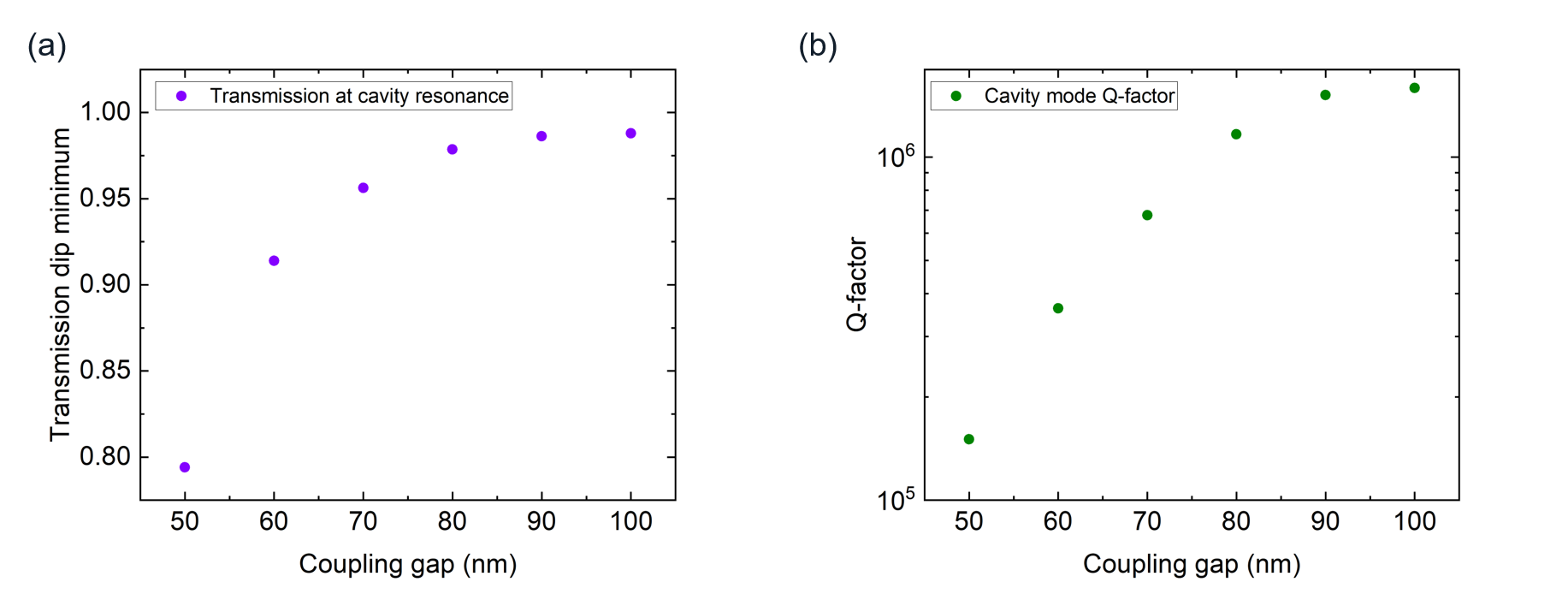}
    \caption{Simulations showing the relationship between the coupling gap and (a) end-to-end transmission through the waveguide (b) cavity mode Q-factor for a fixed bus waveguide width of $w_b = 170$ nm.}
    \label{fig:bwgcouplingd0}
\end{figure}

\autoref{fig:bwgcouplingd0} (a) and (b) show the relationship between the coupling gap and the transmission through the device and cavity mode Q-factor for a fixed bus waveguide width, respectively. As the coupling gap is increased, the magnitude of the transmission dip decreases. This is due to the reduced spatial overlap between the cavity and waveguide modes reducing the coupling efficiency between the two modes. The Q-factor of the cavity mode also increases with increasing coupling gap due to this reduced coupling efficiency. For a fixed waveguide width of 170 nm, the influence of the coupling gap on the transmission dip and cavity Q-factor is totally diminished for $d_0 \geq 100$ nm. 

\newpage
\section{Micro-PL \label{SI:PL} }
\begin{figure}[h!]
\centering
\includegraphics[width=1\textwidth]{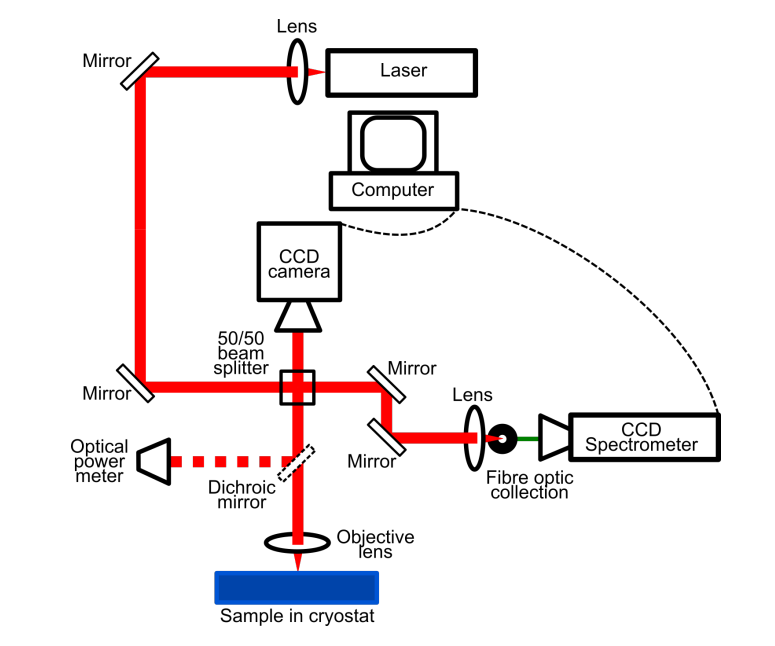}
\caption{Diagram of Micro-PL setup}
\label{fig:MPL}
\end{figure}

\autoref{fig:MPL} illustrates the optical setup used to obtain the results contained in the main text. The chip is placed in a bath cryostat at 4K. A CCD confocal camera is used to image the sample. The QD/cavity emission is collected in the far-field scattering directly over the cavity or OC (refer to main text for each case) travelling through free space out of the device plane to couple optical pump power into the device or propagate emission to a CCD spectrometer or Avalanche single-photon detector
for analysis. All optical pumping was done using a above-band diode laser at 808 nm.

\newpage
\section{Cavity Coupled QD Cross Correlation Measurement \label{SI:HBT}}
The model used to fit the second order cross correlation as a function of time delay between coincidence detection $\tau$ is given by:
\\
\\
\begin{equation*}
    g^{2}(\tau) = 1- (1+a)e^{\frac{-|\tau|}{\tau_1}} + b e^{\frac{-|\tau|}{\tau_2}}
\end{equation*}
\\
\\
Where :
\begin{itemize}
    \item \(\tau_1\) is the anti-bunching characteristic time decay
    \item \(\tau_2\) is the bunching characteristic time decay
    \item \(a,b\) is the scaling of an imperfect system of anti-bunching and bunching respectively
\end{itemize}

Using a model which assuming b=1 we get the following fit:
\\
\\
\begin{table}[h!]
\centering
\begin{tabular}{|l|c|c|}
\hline
\textbf{Parameter} & \textbf{Fitted Value} & \textbf{Error of Fit} \\
\hline
\(\tau_1\) (ns) & 142.261 & $\pm$ 17.350 \\
\(\tau_2\) (ns) & 439.907 & $\pm$ 41.249 \\
\(a\) & 0.872 & $\pm$ 0.135 \\
\(g^{(2)}(0)\) & 0.176 & $\pm$ 0.142 \\
\hline
\end{tabular}
\label{tab:fit_parameters}
\end{table}

\end{document}